\documentclass[runningheads]{llncs}
\usepackage{graphicx}

\usepackage{tikz}
\usepackage{comment}
\usepackage{amsmath,amssymb} 
\usepackage{color}
\usepackage{multirow}
\usepackage{tabularx}

\makeatletter
\def\hlinew#1{%
	\noalign{\ifnum0=`}\fi\hrule \@height #1 \futurelet
	\reserved@a\@xhline}
\makeatother%

\begin{document}
\pagestyle{headings}
\mainmatter
\def\ECCVSubNumber{100}  

\title{Unpaired Deep Image Dehazing Using Contrastive Disentanglement Learning} 


\titlerunning{Unpaired Deep Image Dehazing Using Contrastive Disentanglement Learning}

\author{Xiang Chen\inst{1,2} $^{\dagger}$ \and
Zhentao Fan\inst{1} $^{\dagger}$ \and
Pengpeng Li\inst{3} \and
Longgang Dai\inst{1} \and 
Caihua Kong\inst{1} \and \\
Zhuoran Zheng\inst{2} \and
Yufeng Huang\inst{1} \and
Yufeng Li\inst{1} \textsuperscript{\thanks{Corresponding author. $^{\dagger}$ Authors contributed equally to this work.}}
}

\authorrunning{Chen et al.}
%

\institute{Gaofen Lab, Shenyang Aerospace University
		\and CSE, Nanjing University of Science and Technology 
		\and ISE, Dalian Polytechnic University}
	
\maketitle

\begin{abstract}
We offer a practical unpaired learning based image dehazing network from an unpaired set of clear and hazy images. This paper provides a new perspective to treat image dehazing as a two-class separated factor disentanglement task, \textit{i}.\textit{e}, the task-relevant factor of clear image reconstruction and the task-irrelevant factor of haze-relevant distribution. To achieve the disentanglement of these two-class factors in deep feature space, contrastive learning is introduced into a CycleGAN framework to learn disentangled representations by guiding the generated images to be associated with latent factors. With such formulation, the proposed contrastive disentangled dehazing method (CDD-GAN) employs negative generators to cooperate with the encoder network to update alternately, so as to produce a queue of challenging negative adversaries. Then these negative adversaries are trained end-to-end together with the backbone representation network to enhance the discriminative information and promote factor disentanglement performance by maximizing the adversarial contrastive loss. During the training, we further show that hard negative examples can suppress the task-irrelevant factors and unpaired clear exemples can enhance the task-relevant factors, in order to better facilitate haze removal and help image restoration. Extensive experiments on both synthetic and real-world datasets demonstrate that our method performs favorably against existing unpaired dehazing baselines. 

\end{abstract}

\section{Introduction}
Single image dehazing (SID) is a typical low-level vision problem emerging in recent years, whose aim is to predict the haze-free image from the observed hazy image. Most existing SID methods are immersed in learning supervised models from paired synthetic data \cite{li2018benchmarking}, which inevitably limits their generalization capability in real-world applications. Therefore, learning the practical SID network from an unpaired set of clear and hazy images is significant as obtaining paired real-world data is almost prohibitively expensive and time-consuming \cite{zhao2021refinednet,chen2022unpaired}.

How to learn a SID network when paired data is not available? To solve this issue, some recent studies \cite{engin2018cycle,dudhane2019cdnet,jin2020unsupervised,liu2021synthetic} attempt to explore different unpaired dehazing solutions that mainly divided into two trends. The first one is semi/un-supervised transfer learning \cite{li2019semi,engin2018cycle,dudhane2019cdnet}, where they either utilize the circulatory structure of CycleGAN \cite{zhu2017unpaired} or design domain adaption paradigms \cite{shao2020domain,chang2021damix} to boost the generalization abilities of the algorithm themselves. The above transfer learning based approaches regard SID as an image-to-image translation case, which are performed by making use of the limited labeled data and adding the auxiliary optimization terms. Due to the fact that the domain knowledge of the hazy and haze-free images is asymmetrical, it is laborious for these CycleGAN-based strategies to capture accurate mapping between two different domains using only weak constraints. Furthermore, these methods ignore the potential association in the  latent space \cite{chen2022unpaired} and do not fully mine the useful feature information for SID, resulting in sub-optimal performance.

In consideration of the hazy input as the entanglement of several simple layers (\textit{i}.\textit{e}, the scene radiance layer, the transmission map layer, and the atmospheric light layer), another popular way can be seen as a problem of the physical-based disentanglement. With this idea, several works \cite{yang2018towards,li2020zero,li2021you,liu2021synthetic} fully consider the physical model of haze process, and employ three joint subnetworks to disentangle the given hazy image into these three component parts, so as to estimate the haze and recover the clear image. Although learning disentangled representations has certain natural advantages, it is not easy to disentangle into three hidden factors from the hazy input. Furthermore, since the model is only a rough approximation of the real world, relying on a physics-based model to design the SID network would not make the method robust, especially under non-uniform haze conditions.

Following the above two lines of thinking, we rethink hazy image formation by simplifying the entanglement model itself. Motivated by the similar intuition in \cite{wang2021disentangled,liu2021disentangling}, we make a simple and elegant assumption of factor disentanglement which views an hazy image as an entanglement of two separable parts, a task-relevant factor (\textit{e}.\textit{g}, the color, texture, and semantic information of the clear background image) and a task-irrelevant factor (\textit{e}.\textit{g}, the distribution of the haze component). In this work, our key insight is that a good dehazing model is formulated by enhancing task-relevant factors, while suppressing task-irrelevant factors in the latent space. In other words, it could be helpful to reconstruct a clear image from the learned unambiguous embeddings by clustering these factors with the same value together and isolating other factors with the different value. The intuitive fact is that the same factor values produce the similar image features related to that factor \cite{pan2021contrastive}, and vise versa. Therefore, this encourages us to introduce recent successful contrastive learning into the frequently-used unpaired adversarial framework, CycleGAN, to guide the generated images to be associated with latent factors, so that we can facilitate the learned representation to fulfill factor disentanglement and help image restoration.

In this paper, a contrastive disentangled dehazing method (CDD-GAN) is formulated without using paired training information. Specifically, we introduce a bidirectional disentangled translation network as the backbone of the proposed CDD-GAN. Different from the conventional contrastive loss \cite{park2020contrastive} in GANs, we employ negative generators to perform adversarial contrastive mechanism \cite{wang2021instance} on the image generator encoder, so as to produce a series of challenging negative adversaries. With these hard negative adversaries, the image encoder on the backbone representation network will learn more distinguishing representation of the latent factors, so that we can disentangle the discrete variation of these factors during the bidirectional translation process. When the above-mentioned two-class factors are well separated, the image decoder will better isolate those task-irrelevant factors and obtain a more accurate representation for achieving high-quality outputs. To summarize, we offer the following contributions:
\begin{itemize}
	\item We rethink the image dehazing task and propose an effective unpaired learning framework CDD-GAN, which first attempts to leverage disentangled factor representations to facilitate haze removal in the latent space.
	
	\item We introduce adversarial contrastive loss into CDD-GAN to fulfill factor disentanglement, where hard negative examples can suppress the task-irrelevant factors and unpaired clear exemples can enhance the task-relevant factors.
	
	\item Extensive experiments are carried out on both synthesis and real-world datasets, and demonstrate that our method is superior to existing unpaired dehazing networks and achieves encouraging performance.
\end{itemize}

\section{Related Work}
\subsection{Single image dehazing}
{\bf For the paired dehazing aspect}, many classical methods \cite{cai2016dehazenet,ren2016single,li2017aod,zhang2018densely,liu2019learning} continuously comply with atmospheric scattering model and restore haze-free image through the estimation of the global atmospheric light and transmission map. Nevertheless, these algorithms tend to fail drastically when the corresponding parameter estimation is not accurate enough, thereby resulting in sub-optimal performance. To remedy this, numerous end-to-end dehazing networks \cite{ren2018gated,li2018single,chen2019gated,liu2019griddehazenet,dong2020multi,qin2020ffa} are recently developed for directly outputting dehazed images from hazy inputs without estimating atmospheric lights and transmission maps. However, those paired supervised models in dealing with real-world images will rapidly drop due to the inter-domain and intra-domain gap \cite{yi2021two} between the training and test data.

{\bf For the unpaired dehazing field}, inspired by the popular CycleGAN \cite{zhu2017unpaired}, previous works pursue directly learning the translation relationship from hazy domain to haze-free domain without using paired training information. In \cite{engin2018cycle,dudhane2019cdnet,anvari2020dehaze,liu2020efficient}, several dehazing methods based on improved CycleGAN structure are proposed by utilizing unpaired adversarial learning strategy. Due to the domain knowledge between hazy and clear images is asymmetrical \cite{chang2021damix}, it is not effective to restore high-quality results only relying on limited cycle-consistency constraints. Afterwards, Li et al. \cite{li2019semi} first explore a semi-supervised dehazing framework, which can promote the learning ability by using unlabeled real hazy images amd synthetic images. Recently, the idea of physical-based disentanglement \cite{li2021unsupervised,jin2020unsupervised,zhao2021refinednet} has emerged to further increase the unpaired dehazing performance. For instance, Yang et al. \cite{yang2018towards} design disentangled dehazing network (DisentGAN) to estimate the scene radiance, the medium transmission, and global atmosphere light by exploiting different generators jointly. Similarly, numerous novel unsupervised disentangled network architectures have been developed, such as you only look yourself (YOLY) \cite{li2021you}, zero-shot image dehazing (ZID) \cite{li2020zero}, disentangled-consistency mean-teacher network (DMT-Net) \cite{liu2021synthetic}. Unlike these methods based on complex multilayer disentanglement, our assumption is simpler, that is, the latent space can be further divided into two separated parts, including the task-relevant factors and the task-irrelevant factors.

\subsection{Contrastive learning}
Contrastive loss has demonstrated its effectiveness in self-supervised and unsupervised representation learning \cite{chen2020simple}. Recent researches have employed contrastive learning into low-level vision tasks and obtained improved performance, such as haze removal \cite{wu2021contrastive}, rain removal \cite{chen2022unpaired}, image super-resolution \cite{wang2021unsupervised} and image-to-image translation \cite{park2020contrastive,han2021dual}. The most critical design in contrastive learning is how to select the negatives. Different with previous methods sampling negative examples from patches at different positions in the source image, we actively train a set of negative examples as a whole in an adversarial manner. The closest thing to our method is \cite{wang2021instance}, but the difference is that our method performs the contrastive operation in the CycleGAN framework, which benefits from mining the attributes of unpaired clear exemples in the backward cycle.

\subsection{Disentanglement in GANs}
Disentanglement methods in GANs \cite{chen2016infogan,esser2020disentangling} have been proposed and used to decompose and recombine the representations of individual factors from hidden representations. Most disentanglement frameworks attempt to learn representations which capture different factors of variation in the latent space. Recently, Ye et al. \cite{ye2021closing} decomposed the rainy image into the rain-free background and the rain layer in disentangle image translation framework. Inspired by \cite{pan2021contrastive}, we flexibly embed the contrastive learning into the disentangle translation network to enable the end-to-end training, which could be beneficial to image disentanglement.

\section{Proposed Method}

\subsection{Problem formulation}
Let $\mathcal{D}_{unpair}=\left\{\left(I_{H}^{i}, I_{N}^{\pi_{i}}\right)\right\}_{i=1}^{l}$ be a training dataset for unpaired SID, where the permutation $\pi$ indicates that each pair of the hazy image $I_{H}^{i}$ and the clear image $I_{N}^{\pi_{i}}$ does not have any content correspondence. The goal of unpaired SID is to learn a deep model to explore the intrinsic connection based on the unpaired dataset ${D}_{unpair}$ without the supervision of the ground truth labels to estimate the haze-free images. To achieve the goal, most of the existing disentanglement-based unpaired SID methods empirically construct three joint disentanglement subnetworks under the assumption of atmospheric scattering model. Formally,
\begin{equation}
I_{H}(x)=I_{N}(x)t(x)+A(1-t(x)),
\end{equation}
where $A$ represents the atmospheric light, and $t(x)$ describes the transmission map on each pixel coordinates. Different from these methods that guide the layer disentanglement by describing the hazing process in image space, we rethink hazy image formation by simplifying the entanglement model itself in feature space. From the perspective of feature distribution learning, it can be formulated as
\begin{equation}
p(I_{H})=p(I_{N}, I_{h})=p(I_{N}) p(I_{h} \mid I_{N}),
\end{equation}
where the distribution of the hazy image $p(I_{H})$ is a joint distribution of the clear image $p(I_{N})$ (contains task-relevant factor) and haze component $p(I_{h})$ (contains task-irrelevant factor). The clean representation can be achieved if we can disentangle task-relevant factor $c_{r}$ and task-irrelevant factor $c_{ir}$ from $p(I_{H})$. Then, the clear images can be recovered with the disentangled task-relevant factor.

To achieve factor disentanglement, the recent contrastive representation learning may open a door for guiding the learning of an unambiguous embedding. Due to the intuitive fact is that the same factor values produce the similar image features related to that factor, we propose to compare the features of the generated images to disentangle the discrete variation of these two-class factors. The details of our proposed framework are described below. 

\begin{figure}
	\centering
	\includegraphics[width=0.89\textwidth]{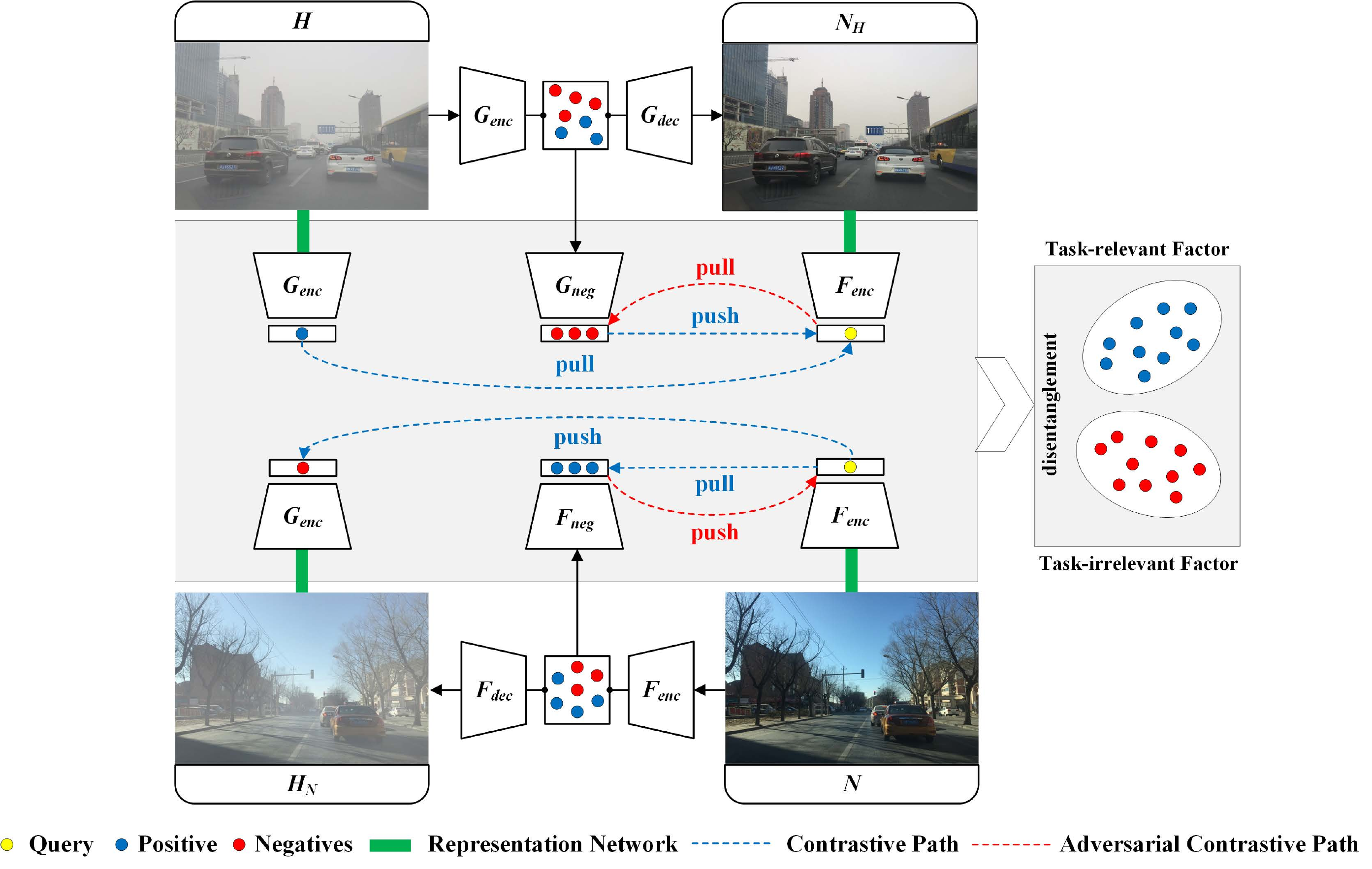}
	\caption{The overview of the proposed contrastive disentangled dehazing method (CDD-GAN). In our framework, two alternately updated paths are playing a minimax game to achieve factor disentanglement in the latent feature space. On one path, the backbone representation network is trained in the conventional contrastive learning. On the other path, the negative generators enforce adversarial contrastive learning to pull negatives to closely track the positive query. Here, we omit two discriminators.}
	\label{fig2}
\end{figure}

\subsection{Framework architecture}
Based on above analysis, we formulate contrastive disentanglement in a GAN framework to achieve better unpaired SID performance. Fig. \ref{fig2} shows the overall architecture of our developed contrastive disentangled dehazing method (CDD-GAN). Since the natural advantages of CycleGAN can fully excavates the useful feature properties of unpaired clear images for SID, we introduce a bidirectional disentangled translation network as the backbone of the proposed CDD-GAN. Intuitively, the first half of the generators are presented as encoders while the second half are decoders, and defined as $G_{enc}$ and $F_{enc}$ followed by $G_{dec}$ and $F_{dec}$ respectively. In our framework, two alternately updated paths (\textit{i}.\textit{e}, contrastive path and adversarial contrastive path) are playing a minimax game to achieve factor disentanglement in the latent space. Ideally, we note that enhancing task-relevant factors as well as isolating task-irrelevant factors will be a double benefit for building a better SID framework \cite{wang2021disentangled}. We will illustrate it with feature visualization in Section 4.4. Thus, the advantage of such a contrastive disentanglement design is twofold. First, the isolation of the task-irrelevant factors can reduce the ambiguity of the encoder network representation to guide the learning of more unambiguous embedding. On the other hand, the enhancement of the task-relevant factors can encourage the capability of the decoder network representation to guide the learning of more accurate mapping.

To capture variability between $p(I_{H})$ and $p(I_{N})$ in feature space, similar to the dual learning setting in \cite{han2021dual}, we first extract features of images from the $L$ layers of $G_{enc}$ and $F_{enc}$, and then send them to a two-layer multi-layer perceptron (MLP) representation network $\mathcal{R}$. Compared with the previous method \cite{chen2022unpaired,park2020contrastive,han2021dual} of randomly sampling negative exemples, we introduce negative generators $G_{neg}$ and $F_{neg}$ to produce more challenging negatives $\mathcal{N}$ based on the embedded features of the image in an adversarial manner, thereby allowing them to closely track the change of representations during the optimization \cite{wang2021instance}. With the help of the adversarial contrastive loss, these hard negative adversaries are trained end-to-end together with the image encoder network (\textit{i}.\textit{e}, the image generator and representation network) to guide the generated images to be associated with latent factors, so that we can facilitate the learned representation to fulfill factor disentanglement during the bidirectional translation process. Finally, the image decoder of CDD-GAN will remove those task-irrelevant factors that are not used to generate the recovered image for generating high-quality dehazed results. The details of the adversarial contrastive loss are illustrated below. 

\subsection{Adversarial contrastive loss}
To disentangle the discrete variation of two-class factors, we define the latent code consists of two parts: category value $c \in \mathcal{Y} =\left\{ c_{r}, c_{ir} \right\}$ and distribution value $z \sim \mathcal{Z}(0,1)$. The generators $G_{enc}$ and $F_{enc}$ take both $c$ and $z$ as inputs and yield generated images $G_{enc}(z,c)$ and $F_{enc}(z,c)$. Take $H \rightarrow N_{H}$ as an example, given a query image $q$ generated from a latent code, we extract feature representations for generated images, \textit{i}.\textit{e}, $f=E_{L}(G_{enc}(z,c))$. We wish the same factor values $c$ produce the similar image features $f$, even match with various $z$, and vice versa \cite{liu2021divco,pan2021contrastive}. Here, we denote the corresponding similar feature as ``positive'' $f^{+}=E_{L}(G_{enc}(z^{+},c^{+}))$ and dissimilar features as ``negatives'' $f_{adv,i}^{-}=E_{L}(G_{enc}(z_{adv,i}^{-},c_{i}^{-}))$. It is worth noting that we actively train a set of negative examples as a whole in an adversarial fashion, and experiments demonstrate our strategy can promote disentanglement performance (see Section 4.4). 

To be specific, conventional contrastive learning is performed to learn a representation for training network backbone, which aims to pull similar feature distribution and push disimilar apart in feature space by minimizing the contrastive loss. On the other hand, we conduct adversarial contrastive learning on the $G_{enc}$ and $F_{enc}$ to cooperate with $\mathcal{R}$ to update alternately by maximizing the contrastive loss. With these negative adversaries produced by the $G_{neg}$ and $F_{neg}$, the encoder network will learn more distinguishing representation of the latent factors, which in turn causes the negative exemples to closely track the positive query. In a word, this leads to a minimax problem, that is, training two mutually interacted players (\textit{i}.\textit{e}, $\mathcal{R}$ and $\mathcal{N}$) jointly with the adversarial contrastive loss $\mathcal{L}_{ac}$ for CDD-GAN. Mathematically, it takes the form:
\begin{equation}
\mathcal{R}^{\star}, \mathcal{N}^{\star}=\arg \min _{\mathcal{R}} \max _{\mathcal{N}} \mathcal{L}_{ac}(\mathcal{R}, \mathcal{N}),
\end{equation}
where $\mathcal{R}$ and $\mathcal{N}$ will reach an equilibrium by alternate training. Generally, a pair of gradient descent and ascent are applied to alternately update the network parameters $\theta_{R}$ and $\theta_{N}$, which are formulated as follow,
\begin{equation}
\theta_{R} \leftarrow \theta_{R}-\eta_{R} \frac{\partial \mathcal{L}_{ac}(\mathcal{R}, \mathcal{N})}{\partial \theta_{R}},
\end{equation}
\begin{equation}
\theta_{N} \leftarrow \theta_{N}+\eta_{N} \frac{\partial \mathcal{L}_{ac}(\mathcal{R}, \mathcal{N})}{\partial \theta_{N}},
\end{equation}
where $\eta_{R}$ and $\eta_{N}$ are the positive learning rates for updating the network and negative adversaries. By constraining the contrastive distribution learning with $\mathcal{L}_{ac}$, these representations become well distinguished and can be formulated as 
\begin{equation}
\mathcal{L}_{ac}= \mathbb{E}_{\mathcal{D}_{unpair}}\left[-\log \frac{\operatorname{sim}\left(f, f^{+}\right)}{\operatorname{sim}\left(f, f^{+}\right)+\sum_{i=1}^{N} \operatorname{sim}\left(f, f_{adv,i}^{-}\right)}\right],
\end{equation}
where $\tau$ is the scalar temperature parameter, and $\operatorname{sim}(u, v)=\exp \left(\frac{u^{T} v}{\|u\|\|v\| \tau}\right)$ is the similarity between the two normalized feature vectors.

\subsection{Other objectives}
Since ground truths are not available, it is essential to constrain CDD-GAN with several effective loss functions. As well as the adversarial contrastive loss mentioned above, we introduce other objectives to regularize the network training process.

{\flushleft \textbf{Diversity loss}}.
To encourage the generation of diverse hard negative exemples $\mathcal{N}_{i}=\left\{N_{0}, N_{1}, \cdots, N_{l}\right\}$ in $\mathcal{L}_{ac}$, similar to \cite{wang2021instance}, we introduce the diversity loss by combining different input noises, which is formulated as follows,
\begin{equation}
\mathcal{L}_{div}=-\left\|\mathcal{N}_{i}\left(\overline{\mathcal{R}}, v_{1}\right)-\mathcal{N}_{i}\left(\overline{\mathcal{R}}, v_{2}\right)\right\|_{1},
\end{equation}
where $\overline{\mathcal{R}}$ denotes the spatially-average features from $\mathcal{R}$, and $v_{i}$ is noise vector randomly sampled from standard Gaussian distribution.
	
{\flushleft \textbf{Total variation loss}}.
To remove the artifacts in the restored images, we apply the total variation to $N_{H}$:
\begin{equation}
\mathcal{L}_{t v}=\left\|\partial_{h} N_{H}\right\|_{1}+\left\|\partial_{v} N_{H}\right\|_{1},
\end{equation}
where $\partial_{h}$ and $\partial_{v}$ represent the horizontal and vertical gradient operators, respectively.

{\flushleft \textbf{Dark channel loss}}. Inspired by \cite{he2010single,li2019semi,shao2020domain}, we also take advantage of the dark channel of clear images, which is written as:
\begin{equation}
D(I)=\min _{y \in N(x)}\left[\min _{c \in\{r, g, b\}} I^{c}(y)\right],
\end{equation}
where $x$ and $y$ are pixel coordinates, $N(x)$ is an image patch centered at $x$, and $I^{c}$ denotes c-th color channel. Thus, we impose dark channel loss to further constrain the sparsity of the dark channel of the dehazed images:
\begin{equation}
\mathcal{L}_{dc}=\left\|D\left(N_{H}\right)\right\|_{1}.
\end{equation}

{\flushleft \textbf{Full objective}}. The full objective function for the negative generator and encoder network are as follows:
\begin{equation}
\mathcal{L}_{neg}=-\mathcal{L}_{ac}+\lambda_{1} \mathcal{L}_{div},
\end{equation}
\begin{equation}
\mathcal{L}_{enc}=\mathcal{L}_{ac}+\lambda_{2} \mathcal{L}_{adv}+\lambda_{3} \mathcal{L}_{cycle}+\lambda_{4} \mathcal{L}_{tv}+\lambda_{5} \mathcal{L}_{dc},
\end{equation}
where $\lambda_{i}$ is balance weight, $\mathcal{L}_{adv}$ and $\mathcal{L}_{cycle}$ are the generative adversarial loss and the cycle-consistency loss. Here, we empirically set $\lambda_{1}$ =  $\lambda_{2}$ = 1, $\lambda_{3}$ = $10^{-1}$, $\lambda_{4}$ = $10^{-3}$, and $\lambda_{5}$ = $10^{-2}$.

\section{Experimental Results}
\subsection{Datasets setup}
{\flushleft \textbf{SOTS and HSTS}}. 
We conduct experiments on a large-scale benchmark dataset, named REalistic Single Image DEhazing (RESIDE) \cite{li2018benchmarking}, which consists of two testing sets, SOTS and HSTS. In detail, SOTS has 500 indoor and outdoor hazy images generated using the physical model with manual parameters. HSTS provides a synthetic set and a real-world set, each containing 10 hazy images.

{\flushleft \textbf{Foggy Cityscapes}}. 
Sakaridis et al. \cite{SDV18} apply fog simulation on the Cityscapes dataset \cite{cordts2016cityscapes} and generate Foggy Cityscapes with 20,550 images. Here, we select elaborately 4,000 high-quality synthetic hazy-clear images, which contains 3,600 hazy images for training and the remaining 400 ones for evaluation.

\subsection{Training details}
The developed CDD-GAN is based on CycleGAN \cite{zhu2017unpaired}, a Resnet-based generator with nine residual blocks and a PatchGAN \cite{isola2017image} discriminator. The whole framework is implemented using the PyTorch with two Tesla V100 GPUs. We perform the adversarial contrastive learning on the 1-st, 5-th, 9-th, 13-th, 17-th layers of $G_{enc}$ and $F_{enc}$. The number of negative exemples $\mathcal{N}_{i}$ for contrastive learning is set to 256. The temperature parameter $\tau$ is set to 0.07. We apply the Adam optimizer and the batch size is set to 1 and the models are trained for total 400 epochs. Initially, the proposed network is trained with 0.0001 learning rate for 200 epochs, followed by another 200 epochs with linearly decaying learning rate. $256 \times 256$ patches are randomly cropped from all training images in an unpaired learning procedure.

\subsection{Comparison results}
We compare our method with those of two prior-based approaches (i.e., DCP \cite{he2010single} and CAP \cite{zhu2015fast}), three paired learning-based models (i.e., MSCNN \cite{ren2016single}, AODNet \cite{li2017aod}, and GFN \cite{ren2018gated}), four unpaired learning-based networks (i.e., CycleGAN \cite{zhu2017unpaired}, DisentGAN \cite{yang2018towards}, SSID \cite{li2019semi}, and RefineDNet \cite{zhao2021refinednet}). With the help of the corresponding labels in synthetic datasets, we adopt two evaluation criteria: PSNR and SSIM \cite{wang2004image}. To compare real-world hazy cases that lack ground truth, we use the no-reference quality metric NIQE \cite{mittal2012no}.

\begin{table*}[t]\small
	\centering
	\caption{Comparison of quantitative results on three synthetic datasets. Bold and \underline{underline} indicate the best and second-best results.}
	\begin{tabular}{cc|cc|cc|cc}
	\hlinew{1.0pt}
	\multicolumn{2}{c|}{Datasets}                                                                                                                & \multicolumn{2}{c|}{SOTS}       & \multicolumn{2}{c|}{HSTS}       & \multicolumn{2}{c}{Cityscapes} \\ \hline
	\multicolumn{2}{c|}{Metrics}                                                                                                                 & PSNR           & SSIM           & PSNR           & SSIM           & PSNR              & SSIM             \\ \hline
	\multicolumn{1}{c|}{\multirow{2}{*}{Prior-based methods}}                                                                    & DCP \cite{he2010single}          & 16.62          & 0.817          & 14.84          & 0.761          & 15.09             & 0.795            \\
	\multicolumn{1}{c|}{}                                                                                                        & CAP \cite{zhu2015fast}          & 19.05          & 0.836          & 21.53          & 0.872          & 17.34             & 0.844            \\ \hline
	\multicolumn{1}{c|}{\multirow{3}{*}{\begin{tabular}[c]{@{}c@{}}Paired / Supervised\\ methods\end{tabular}}}                  & MSCNN \cite{ren2016single}        & 17.57          & 0.810          & 18.64          & 0.817          & 17.98             & 0.828            \\
	\multicolumn{1}{c|}{}                                                                                                        & AOD-Net \cite{li2017aod}      & 19.06          & 0.850          & 20.55          & 0.897          & 18.51             & 0.836            \\
	\multicolumn{1}{c|}{}                                                                                                        & GFN \cite{ren2018gated}          & 22.30          & 0.884 & \underline{21.87} & 0.893          & 19.69    & 0.857   \\ \hline
	\multicolumn{1}{c|}{\multirow{5}{*}{\begin{tabular}[c]{@{}c@{}}Unpaired / Without paired\\ supervised methods\end{tabular}}} & CycleGAN \cite{zhu2017unpaired}     & 17.78          & 0.725          & 18.52          & 0.831          & 17.82             & 0.812            \\
	\multicolumn{1}{c|}{}                                                                                                        & DisentGAN \cite{yang2018towards}    & 22.12          & 0.899          & 19.68              & 0.866              & 18.66                 & 0.837                \\
	\multicolumn{1}{c|}{}                                                                                                        & SSID \cite{li2019semi}         & \underline{24.44}          & 0.896          & 21.83          & 0.882          & 19.50             & 0.841            \\
	\multicolumn{1}{c|}{}                                                                                                        & RefineDNet \cite{zhao2021refinednet}   & 24.39          & \underline{0.912}          & 21.69          & \underline{0.904}          & \underline{20.24}            & \underline{0.866}            \\
	\multicolumn{1}{c|}{}                                                                                                        & \textbf{Ours} & \textbf{24.61} & \textbf{0.918}          & \textbf{22.16} & \textbf{0.911} & \textbf{20.93}    & \textbf{0.874}   \\ \hlinew{1.0pt}
\end{tabular}
\label{table1}
\end{table*}

\begin{figure}
	\centering
	\includegraphics[width=1.0\textwidth]{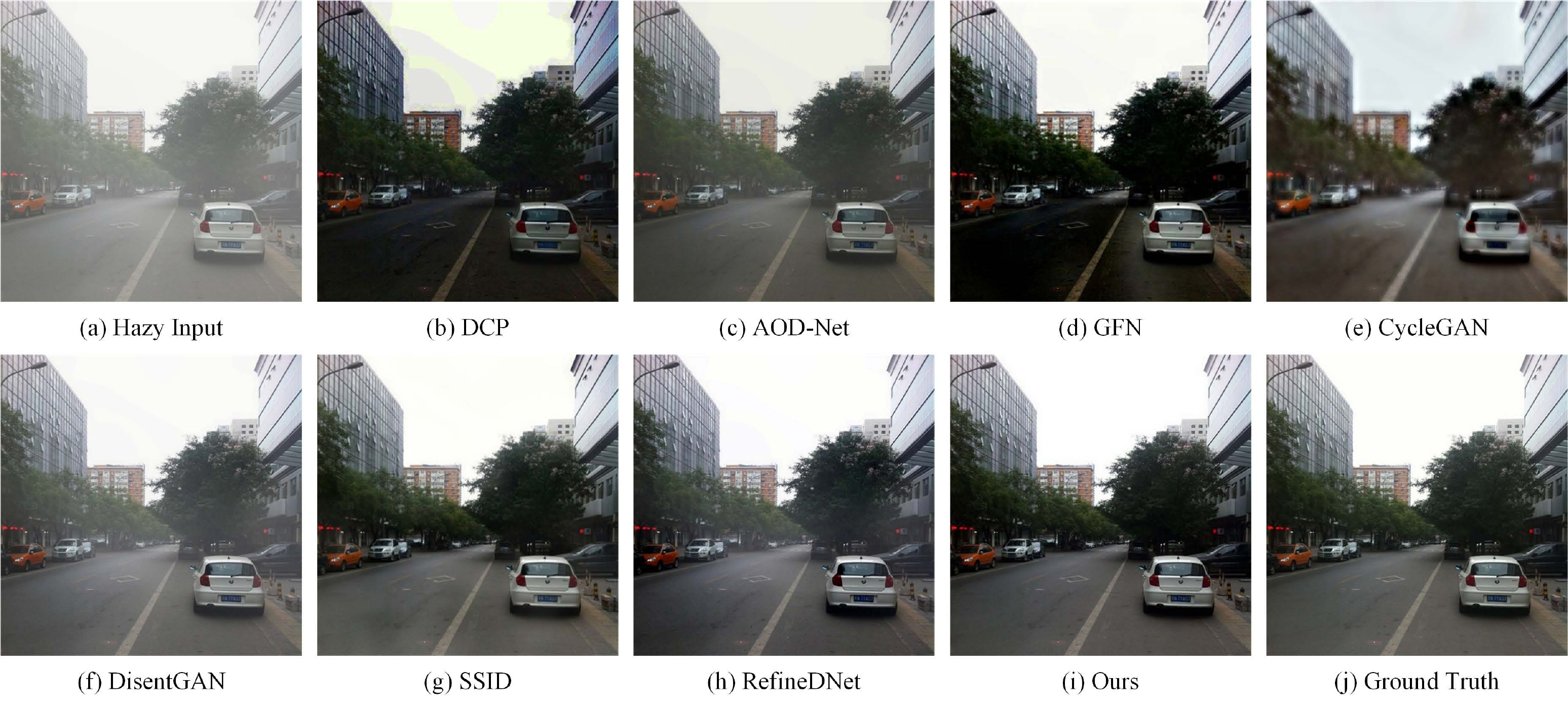}
	\caption{Comparison of qualitative results on the SOTS synthetic dataset.}
	\label{fig3}
\end{figure}

\begin{figure}
	\centering
	\includegraphics[width=1.0\textwidth]{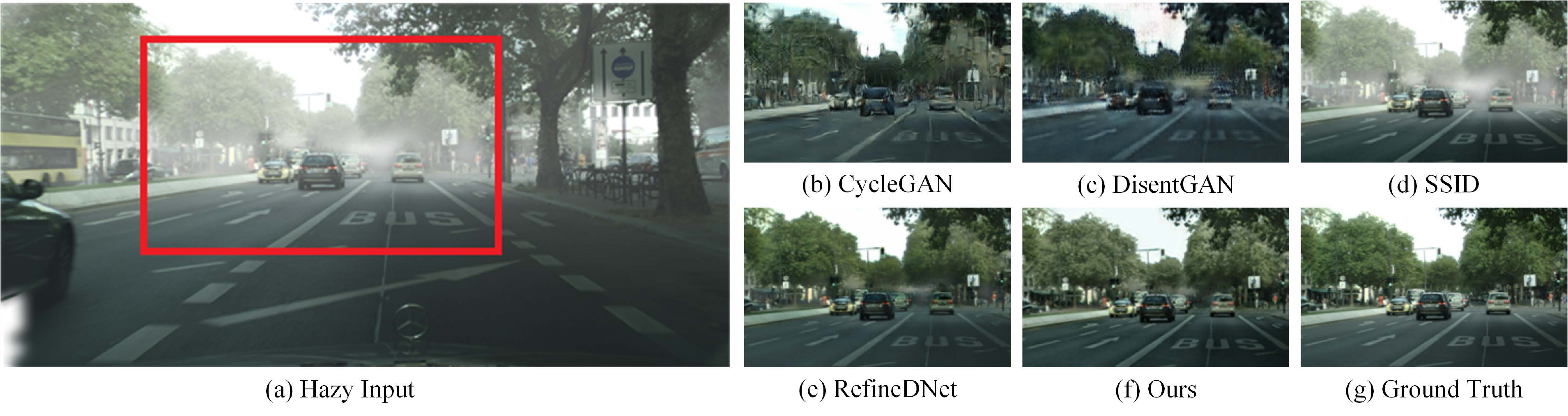}
	\caption{Comparison of qualitative results on the Foggy Cityscapes synthetic dataset.}
	\label{fig4}
\end{figure}

{\flushleft \textbf{Results on synthetic datasets}}. Table \ref{table1} summarizes quantitative values of different approaches on synthetic datasets including SOTS, HSTS, and Foggy Cityscapes. We can notice that our method remarkably outperforms all existing unpaired dehazing nets and achieves state-of-the-art performance. Despite the unsupervised characteristics of our proposed CDD-GAN, it can also deliver comparable results against several paired supervised models, clearly demonstrating that the potential advantages of our proposed contrastive disentanglement framework. Besides the quantitative results, we further present visual observation comparisons in Fig. \ref{fig3} and Fig. \ref{fig4}. All the competitive methods contain more haze residue and obtain unsatisfactory results in detail restoration, which keep consistent with the above quantitative scores. In contrast, the proposed method generates much clearer results that are visually close to the ground truth.

\begin{figure}
	\centering
	\includegraphics[width=1.0\textwidth]{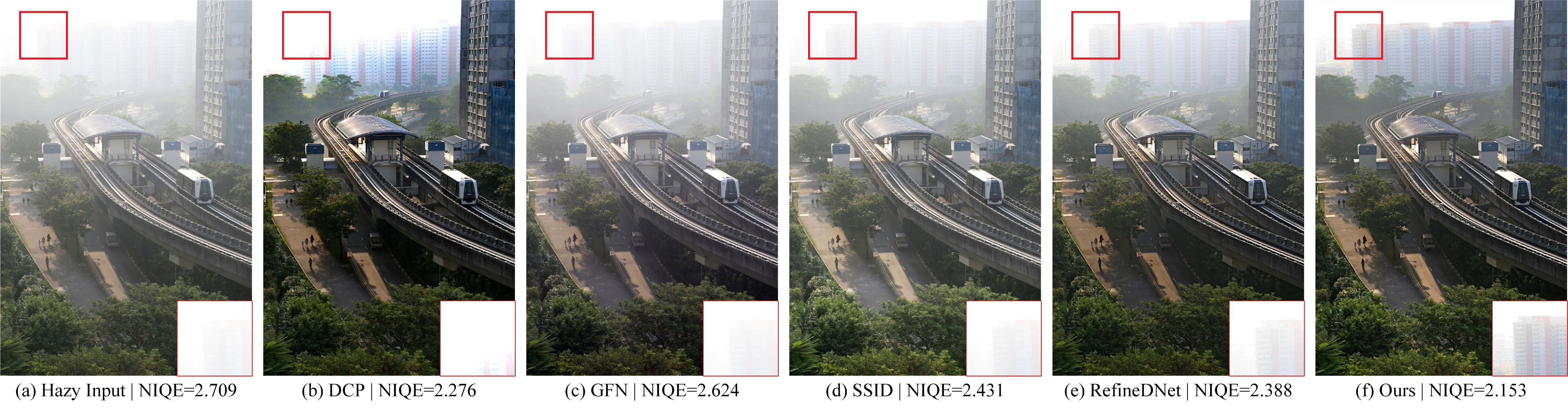}
	\caption{Comparison of qualitative and quantitative results on the HSTS real-world set. Note that lower values of NIQE indicate better image quality.}
	\label{fig5}
\end{figure}

{\flushleft \textbf{Results on real-world datasets}}. To demonstrate the effectiveness of our dehazing model on real hazy images, we conduct comparisons against other algorithms on the HSTS real-world image set and present results in Fig. \ref{fig5}. According the values of NIQE under the images, the proposed method obtains the lowest score, which indicates a high-quality dehazed result with better fidelity and higher naturalness. This benefits from the fact that the decoder of CDD-GAN can reconstruct high-quality outputs with the help of factor disentanglement. 

\subsection{Ablation analysis and discussion}
We study the main component impacts and parameter choices on the final performance. To ensure the fair comparison, all the ablation studies are performed in the same environment and training settings using the Foggy Cityscapes dataset.

\begin{figure}
	\centering
	\includegraphics[width=1.0\textwidth]{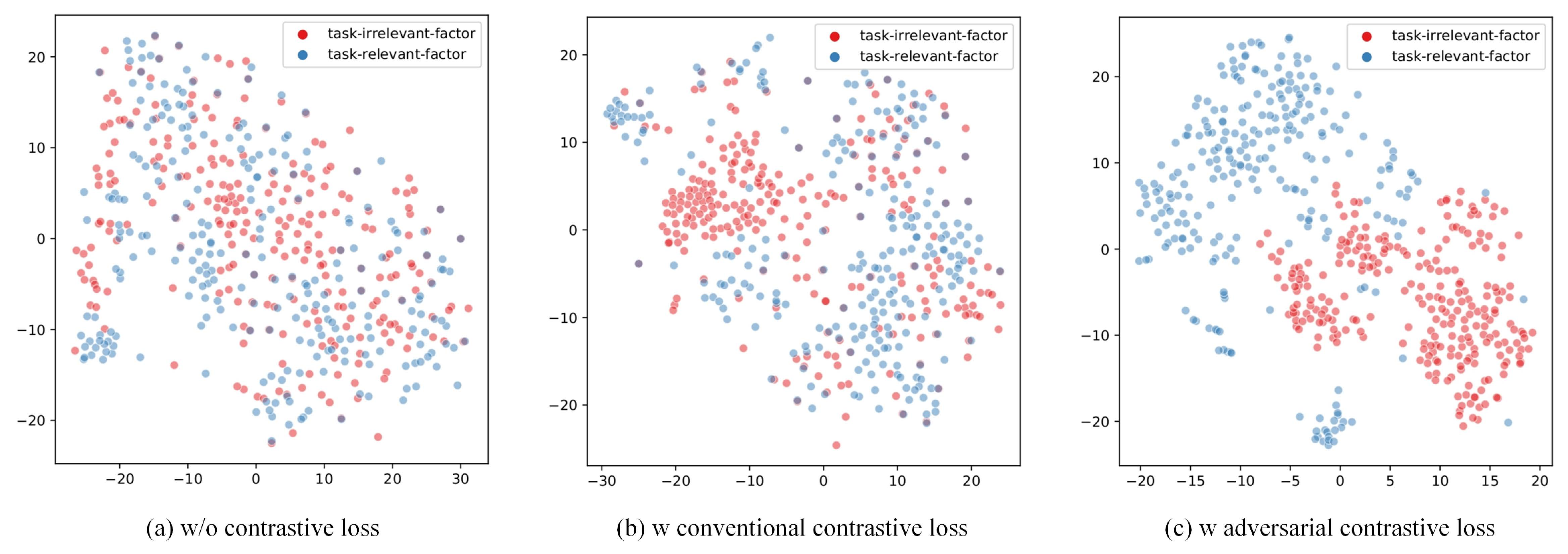}
	\caption{The t-SNE visualization of features learned for task-relevant (blue round point) and task-irrelevant (red round point) factors. With the adversarial contrastive loss, the same factors are pulled closer and gathered together in the latent space. Hence, isolating the task-irrelevant factors by disentanglement is able to generate clear images.}
	\label{fig6}
\end{figure}

{\flushleft \textbf{Effectiveness of negative generator}}.
To investigate the impact of the proposed negative generator, we consider two variants of our framework, including (a) without contrastive loss, and (b) the developed adversarial contrastive loss $\mathcal{L}_{ac}$ is replaced by conventional contrastive loss $\mathcal{L}_{con}$ in \cite{park2020contrastive,han2021dual}. Table \ref{table2} reports the quantitative results of different models. Obviously, contrastive learning can bring great performance gain to the baseline model (a), which shows its potential in unsupervised vision tasks. By comparing model (b) and model (c) in Table \ref{table2}, it reveals that the design of negative generator is more effective than the previous strategy of generating negatives by randomly sampling from the images. To better understand the influence of the negative adversaries in $\mathcal{L}_{ac}$, we further use t-SNE \cite{van2008visualizing} to visualize learned features in Fig. \ref{fig6}. As can be seen, conventional contrastive method is not enough to fulfill factor disentanglement, because their negatives are not effective to push the positives close to the query examples. By contrast, our strategy can produce more discriminative representations, thanks to the challenging negative adversaries provided by negative generators. In such case, the task-irrelevant factors are suppressed by generating hard negative exemples to guide the dehazing process and clear image reconstruction.

\begin{table*}[t]
	\centering
	\caption{Ablation study for different components and designs. PSNR and SSIM results among different models of CDD-GAN on the Foggy Cityscapes dataset. Note that $\mathcal{L}_{con}$  indicates general contrastive loss \cite{park2020contrastive} and  ${S}_{dual}$ indicates dual learning setting.}
    \setlength{\tabcolsep}{2mm}
	\begin{tabular}{c|cccccccc|c}
	\hlinew{1.0pt}
	Models & $\mathcal{L}_{ac}$ & $\mathcal{L}_{con}$ & $\mathcal{L}_{div}$ & $\mathcal{L}_{tv}$ & $\mathcal{L}_{dc}$ & $\mathcal{L}_{adv}$ & $\mathcal{L}_{cycle}$ & ${S}_{dual}$ & PSNR / SSIM   \\ \hline
	(a) & $\times$  & $\times$   & $\times$   & $\times$   & $\times$   & $\checkmark$  & $\checkmark$  & $\times$ & 18.64 / 0.830 \\
	(b) & $\times$  & $\checkmark$   & $\times$   & $\times$   & $\times$   & $\checkmark$  & $\checkmark$  & $\checkmark$ & 19.88 / 0.846 \\
	(c) & $\checkmark$  & $\times$  & $\times$   & $\times$   & $\times$   & $\checkmark$  & $\checkmark$  & $\checkmark$ & 20.35 / 0.843 \\
	(d) & $\checkmark$  & $\times$  & $\checkmark$  & $\times$   & $\times$   & $\checkmark$  & $\checkmark$  & $\checkmark$ & 20.71 / 0.865 \\
	(e) & $\checkmark$  & $\times$  & $\checkmark$  & $\checkmark$  & $\times$   & $\checkmark$  & $\checkmark$  & $\checkmark$ & 20.79 / 0.871 \\
	(f) & $\checkmark$  & $\times$  & $\checkmark$  & $\checkmark$  & $\checkmark$  & $\checkmark$  & $\checkmark$  & $\checkmark$ & \textbf{20.93} / \textbf{0.874} \\
	(g) & $\checkmark$  & $\times$  & $\checkmark$  & $\checkmark$  & $\checkmark$  & $\checkmark$  & $\checkmark$ & $\times$  & 20.72 / 0.863 \\ \hlinew{1.0pt}
\end{tabular}
	\label{table2}
\end{table*}

{\flushleft \textbf{Effectiveness of other objectives}}.
To better demonstrate the effectiveness of other objective function, we also conduct an ablation study by considering the combinations of the diversity loss $\mathcal{L}_{div}$, total variation loss $\mathcal{L}_{tv}$, and dark channel loss $\mathcal{L}_{dc}$. Since our framework is based on CycleGAN, the generative adversarial loss $\mathcal{L}_{adv}$ and cycle-consistency loss $\mathcal{L}_{cycle}$ are the default common items, which will not be discussed here. Correspondingly, we regularly add one component to each configuration at one time. By comparing model (c) and model (d) in Table \ref{table2}, there is a sharp decline in performance without $\mathcal{L}_{div}$, especially on the metric of PSNR. This is because the lack of $\mathcal{L}_{div}$ will cause the negative exemples to maintain less diversity in the training stage. Under this case, the negative generator fails to produce challenging negative exemples, resulting in suboptimal performance. With the combination of all objectives, our model (f) can achieve the best performance, which also demonstrates that each loss term contributes in its own way during dehazing process.

\begin{table*}[t]\small
	\centering
	\caption{Ablation study for different number of negative exemples. PSNR and SSIM results among different settings of CDD-GAN on the Foggy Cityscapes dataset.}
	\setlength{\tabcolsep}{2mm}
	\begin{tabular}{c|cccc}
		\hlinew{1pt}
		Number of $\mathcal{N}_{i}$ & \textit{N} = 64          & \textit{N} = 128         & \textit{N} = 256  (default)                 & \textit{N} = 512         \\ \hline
		PSNR / SSIM    & 19.84 / 0.859 & 20.52 / 0.868 & \textbf{20.93} / 0.874  & 20.85 / \textbf{0.876} \\ \hlinew{1pt}
	\end{tabular}
	\label{table3}
\end{table*}

\begin{figure}
	\centering
	\includegraphics[width=1.0\textwidth]{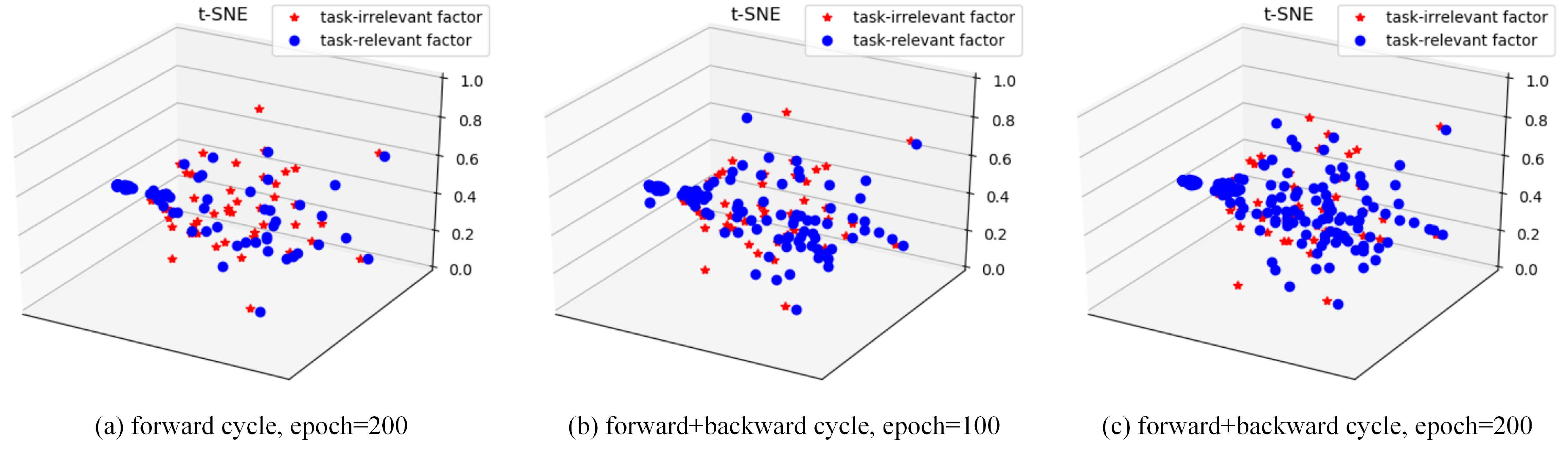}
	\caption{The t-SNE visualization of features learned for task-relevant (blue round point) and task-irrelevant (red pentagram) factors. With the forward and backward dual-path cycle, more task-relevant factors are produced during the optimization. Thus, removing haze can be facilitated by utilizing the features from the unpaired clear exemplars.}
	\label{fig7}
\end{figure}

{\flushleft \textbf{Effectiveness of dual setting}}.
We remove the dual setting ${S}_{dual}$ for comparison, see model (f) and model (g) in Table \ref{table2}. It can be seen that ${S}_{dual}$ achieves dehazing performance improvement, due to its ability to stabilize the training and learn better embeddings for different domains.

{\flushleft \textbf{Influences of unpaired clear exemples}}.
Unlike \cite{park2020contrastive,wang2021instance}, we extend unidirectional mapping to bidirectional mapping, which is more suitable for SID task because it can take advantage of the characteristics of backward cycle. We visualize the corresponding features using t-SNE \cite{van2008visualizing} in Fig. \ref{fig7}. It can be observed that the features from unpaired clear exemples generated by the backward cycle can continuously enhance the task-relevant factors during the optimization.

{\flushleft \textbf{Number of negative exemples}}.
We study the different number influences of negative exemples in Table \ref{table3}. For one thing, too few negatives may weaken the ability to pull the positives closer to the query. For another, too many negatives may increase the computation cost and produce unnecessary interference. To balance the model performances and memory, we choose $\textit{N}$ = 256 as the default.

\subsection{Other applications}
\begin{figure}
	\centering
	\includegraphics[width=1.0\textwidth]{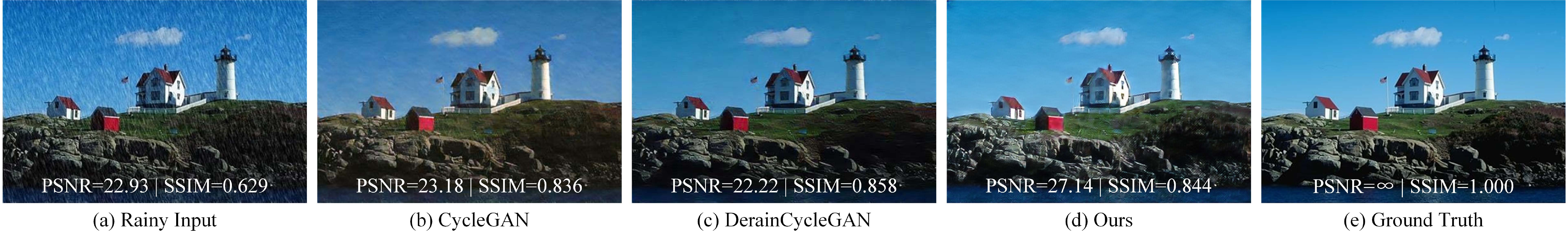}
	\caption{Comparison of qualitative and quantitative results on the Rain800 synthetic dataset \cite{zhang2019image}.}
	\label{fig8}
\end{figure}

{\flushleft \textbf{Generality to other low-level vision tasks}}.
It is a general assumption to regard the degraded image as the entanglement of task-relevant factor and task-irrelevant factor, so our method can be easily applied to similar vision tasks, such as image denoising and deraining. Here, we provide one deraining example in Fig. \ref{fig8} for comparison. Surprisingly, our model even outperforms recent unsupervised deraining method \cite{wei2021deraincyclegan} with more than 5 dB in PSNR. This is because, previous approaches learn complex mapping in the high-dimensional image space, while our method learns latent restoration in the low-dimensional feature space.

\begin{table*}[t]
	\centering
	\caption{Comparison of object detection quantitative results on the RTTS dataset.}
	\setlength{\tabcolsep}{2mm}
	\begin{tabular}{c|cccccc}
	\hlinew{1pt}
	& Hazy Input & CycleGAN & DisentGAN & SSID  & RefineDNet & Ours   \\ \hline
	mAP(\%) & 61.35      & 53.82    & 63.59     & 61.74 & 65.22      & \textbf{66.04} \\ 
	Gain    & —          & -7.53    & +2.24     & +0.39 & +3.87      & \textbf{+4.69}  \\ \hlinew{1pt}
\end{tabular}
	\label{table4}
\end{table*}

{\flushleft \textbf{Preprocessing for high-level vision tasks}}.
As suggested in \cite{yang2020advancing,vidalmata2020bridging,chen2021psd}, SID has become a frequently-used preprocessing step, so we further examine whether our method bring benefits to downstream high-level vision tasks. Here, we randomly select 100 real hazy images from RTTS \cite{li2018benchmarking} with bounding boxes and object categories. We adopt YOLOv3 \cite{redmon2018yolov3} to evaluate object detection performance and then calculate the mean Average Precision (mAP). As can be seen from Table \ref{table4}, the quantitative gain of CycleGAN is negative. This is because it is not completely developed for SID task, which leads to the destruction of the semantic information of the original image. Compared with other unsupervised models, our dehazed results bring higher recognition accuracy for object detection, which further demonstrates the effectiveness of our designed CDD-GAN.

\subsection{Limitations}
In this study, we have three limitations: 1) For $\mathcal{L}_{ac}$, no theory can guarantee convergence to the saddle point, so our method only achieves an approximate equilibrium by updating $\mathcal{R}$ and $\mathcal{N}$ alternately. 2) Due to the common inference of GAN and contrastive learning, the network training is not stable. 3) Our method fails to deal with heavy fog scenes.

\section{Conclusions}
This paper provides a new contrastive disentangled dehazing method (CDD-GAN) to address the challenging unpaired SID problem. To fulfill factor disentanglement in latent feature space, negative generators are introduced into the CycleGAN framework to isolate task-irrelevant factors with the help of adversarial contrastive loss. On the other hand, the features from unpaired clear exemples are utilized for enhancing the learning of task-relevant factors. Extensive experiments considerably show that the effectiveness and scalability of our model. In future work, we plan to explore the possibility of applying contrastive learning for more complex factor disentangling problem in the the field of low-level vision.

\clearpage
%
%
\bibliographystyle{splncs04}
\bibliography{egbib}
\end{document}